\documentclass{ws-procs9x6}            

\newcommand{\ket}[1]{| \, #1 \, \rangle}

\begin{document}
\title{Dynamically generated hadron resonances}

\author{Tetsuo Hyodo}

\address{Department of Physics, Tokyo Metropolitan University, \\
Hachioji 192-0397, Japan \\
E-mail: hyodo@tmu.ac.jp}

\begin{abstract}
Recent developments in hadron spectroscopy triggers the discussion on various exotic configurations beyond the simple three-quark state for baryons and quark-anti-quark pair for mesons. In particular, the states observed near a two-hadron threshold are often regarded as candidates of hadronic molecules, which are made of constituent hadrons and dynamically generated by the hadronic interactions. However, identification of the hadronic molecules involves several subtle difficulties which are often overlooked or underestimated. Here we summarize these subtleties and present a promising approach to distinguish the dynamically generated states from others using observable quantities.

\end{abstract}


\bodymatter

\section{Introduction}

In the history of hadron spectroscopy, more than 300 species of hadrons have been reported to exist.\cite{Tanabashi:2018oca} While all these structures should eventually stem from QCD, we still do not fully understand how they are constructed. For instance, the flavor quantum numbers of the observed hadrons exhibit an empirical regularity; the isospin and strangeness of baryons $(I_{B},S_{B})$ are restricted to
\begin{align}
   I_{B}=0 , \ \frac{1}{2}, \ 1,\ \frac{3}{2},\quad
   S_{B}=0,\ -1,\ -2,\ -3 ,
\end{align}
and the isospin and strangeness of mesons $(I_{M},S_{M})$ have values of
\begin{align}
   I_{M}=0, \ \frac{1}{2},\ 1,\quad
   S_{M}=0,\ \pm \frac{1}{2},\ \pm 1 .
\end{align}
These are exactly the quantum numbers which can be achieved by three valence quarks ($I_{B},S_{B}$), and one quark and one antiquark ($I_{M},S_{M}$). Corresponding flavor SU(3) representations $(\bm{1},\ \bm{3},\ \bar{\bm{3}},\ \bm{6},\ \bm{8},\ \bm{10}$ for baryons and $\bm{1},\ \bm{3},\ \bar{\bm{3}},\ \bm{8}$ for mesons) are classified by ``exoticness'' quantum numbers.\cite{Kopeliovich:1990ez,Diakonov:2003ei,Jenkins:2004tm,Hyodo:2006yk,Hyodo:2006kg} There have been no established hadrons having the flavor quantum numbers other than those given above. This is the ``exotic hadron problem'', for which any reasonable explanation from QCD has not been made.

Recent developments in spectroscopy of hadrons indicate a bunch of candidates of unconventional states, which are considered to have exotic configurations such as multiquark states, gluon hybrids, and hadronic molecules. Here, we do not deal with the configurations characterized by quark-gluon degrees of freedom, but rather focus on hadronic molecules\cite{Guo:2017jvc} which are composite objects of multiple hadrons, because of the following reasons. First of all, the number of constituent quarks and gluons is not a conserved quantum number in QCD (gluons and quark-antiquark pairs can be created or annihilated). Of course, the quark number (the number of quarks minus number of antiquarks) is conserved, but it merely gives the distinction between mesons, baryons, dibaryons, and so on. In other words, there is no well-defined ``multiquark states'' in QCD (except for the characterization by exoticness using flavor quantum numbers, or $J^{PC}$ exotics using spin, parity, and charge conjugation symmetries). On the other hand, while the number of hadrons are not conserved (mesons can be created or annihilated), there is a way to distinguish the two-hadron component from others because hadrons are the asymptotic states in QCD.\cite{Hanhart:2007cm,Hyodo:2013nka} In addition, near-threshold $s$-wave bound states have a spatially large wave function irrespective to microscopic details, thanks to the low-energy universality.\cite{Braaten:2004rn,Naidon:2016dpf} These facts imply that the structure of hadrons should be characterized by the hadronic degrees of freedom. However, there are still subtleties in identifying the hadronic molecules in the spectrum of hadrons. 

\subsection{Hadron resonances}

We first note that most hadrons are unstable states, which eventually decay into two or more hadrons through the strong interaction. Among a few hundred of hadrons listed in Ref.~\citenum{Tanabashi:2018oca}, stable hadrons occupy only small fraction, the ground states in each sector. All the other hadrons, in particular the candidates of exotic hadrons, are located above two-hadron threshold. This is partly because of the light mass of pions, the Nambu-Goldstone bosons of chiral symmetry, in comparison with the internal excitation energy of hadrons. The unstable hadrons should be treated as resonances in hadron scatterings. Hence, the structure of hadrons should be discussed with keeping their unstable nature in mind.

In quantum mechanics, an unstable resonance state shares common features with a stable bound state. For instance, both are represented by an eigenstate of the Hamiltonian with the outgoing boundary condition,\cite{Gamow:1928zz,Siegert:1939zz,Moiseyev} which corresponds to the pole of the S matrix in scattering theory.\cite{Taylor} This observation encourages us to discuss the structure of a resonance by regarding it as a ``state''. At the same time, however, there are also some important differences. The resonance wave function is not square integrable, and its magnitude diverges at large distance. The diverging wave function is physically understood as a consequence of the accumulation of the decaying component at large distance. Formally, the non-square-integrable wave function is required to accommodate the complex eigenvalue of the Hamiltonian, but it also leads to complex expectation values for any other operators. Namely, the properties of a resonance state (norm of each component, mean squared radius, etc.) are obtained as complex numbers, whose interpretation is not straightforward.

\subsection{Dynamically generated states}\label{sec:dg}

Usually, a state is regarded as a dynamically generated one, if it appears in the physical spectrum, in addition to the states prepared in the original model space. To formulate this statement in quantum mechanics, we regard the physical spectrum as the eigenstates of the full Hamiltonian $H$, and the original model space as the eigenstates of the free Hamiltonian $H_{0}$ which is defined as $H=H_{0}+V$ with the interaction $V$. The dynamically generated state is identified as the one which exists only in the eigenstates of $H$. This classification is well defined if certain decomposition of the Hamiltonian $H_{0}+V$ is prepared in advance. In hadron physics, however, the Hamiltonian at the level of hadrons are not given a priori. More specifically, while the full Hamiltonian $H$ is determined by experiments, its decomposition into the free part $H_{0}$ and the interaction $V$ is not unique. In fact, it is explicitly demonstrated by Weinberg\cite{Weinberg:1962hj} that one can add new degrees of freedom in the model space (eigenstates of $H_{0}$) without changing the observables (eigenstates of $H$), by modifying the interaction $V$ appropriately. This is essentially a unitary transformation of the potential which leaves the on-shell quantities invariant. In this way, there is an ambiguity in identifying the dynamically generated states with respect to the model space.

\subsection{Compositeness}

A solution to the problem in the previous section was given also by Weinberg.\cite{Weinberg:1965zz} It is shown that the ambiguity of the model space can be well controlled when we deal with an weakly bound $s$-wave state. By establishing the weak-binding relation, the deuteron is shown to be a composite system of a proton and a neutron, as we describe below. In that work, the structure of the bound state is characterized by the wave function renormalization constant of the bare (elementary) state. Its counterpart, which represents the two-body composite component, is now called compositeness. It is recently shown that the compositeness can be evaluated from the residue of the pole of the scattering amplitude,\cite{Hyodo:2011qc,Aceti:2012dd} and the compositeness of several hadron resonances have been calculated (See Ref.~\citenum{Hyodo:2013nka} and references therein). 

Although the weak-binding relation is a nice tool to investigate the internal structure, with its original derivation, applicability was limited to stable bound states. It is, therefore, necessary to generalize the weak-binding relation to unstable resonances, for the application to hadron physics. This direction of study has been initiated in Ref.~\citenum{Baru:2003qq}. Eventually, direct generalization of the weak-binding relation for unstable resonances is developed with the effective field theory.\cite{Kamiya:2015aea,Kamiya:2016oao}

In the following, we briefly revisit the original weak-binding relation for stable bound states from the viewpoint of effective field theory. Next, we present the generalized weak-binding relation for unstable resonances based on Refs.~\citenum{Kamiya:2015aea,Kamiya:2016oao}. Finally, we show an application of the generalized relation to $\Lambda(1405)$ resonance.

\section{Which hadron is dynamically generated?}

\subsection{Weak-binding relation}

We first consider an $s$-wave stable bound state in a single-channel scattering. For concreteness, let us take an example of the deuteron in the $NN$ scattering. Weinberg's weak-binding relation is given by\cite{Weinberg:1965zz}
\begin{align}
   a_0=R\left\{\frac{2X}{1+X}+\mathcal{O}\left(\tfrac{R_{\rm typ}}{R}\right)\right\},\quad R=\frac{1}{\sqrt{2\mu B}}
   \label{eq:weakbinding} ,
\end{align}
where $a_{0}$ is the scattering length which governs the low-energy $NN$ scattering, and $R$ is the ``radius'', the length scale associated with the binding energy $B$ with the reduced mass $\mu$. $R_{\rm typ}$ is the typical length scale of the two-body interaction, which is estimated by the mass of the lightest possible exchanged particle. For the deuteron, it is given by the pion mass as $R_{\rm typ}\sim 1/m_{\pi}$. The compositeness $X$ is defined as the weight of the two-body component of the wave function, namely, 
\begin{align}
   \ket{d} = \sqrt{X}\ket{NN} + \sqrt{1-X}\ket{\rm others} .
\end{align}
Using Eq.~\eqref{eq:weakbinding}, we can determine the compositeness $X$ from $a_{0}$ and $B$, if the correction term of order $\mathcal{O}\left(R_{\rm typ}/R\right)$ is sufficiently small. In other words, if the binding energy is small and the condition $R\gg R_{\rm typ}$ is met, then the internal structure of the bound state ($X$) is reflected in the observable quantities ($a_{0},B$). By using the experimental values of $a_{0}$ and $B$, it is concluded that the deuteron is dominated by the $NN$ composite system.\cite{Weinberg:1965zz}

Originally, the quantity $Z=1-X$ is introduced as the wave function renormalization constant of a bare state, and interpreted as the weight of an elementary component. From the modern viewpoint, it is the compositeness $X$ that is determined by the weak-binding relation, and  $1-X$ represents the weight of all the components other than the two-body composite one. The weak-binding relation~\eqref{eq:weakbinding} can be derived in the effective field theory with the two-body continuum states coupled with a single discrete state through the contact interactions.\cite{Kamiya:2015aea,Kamiya:2016oao} This formulation provides a clear interpretation of the weak-binding relation with respect to the renormalization. Another virtue of the effective field theory is that it enables us to generalize the weak-binding relation for unstable quasi-bound states.

\subsection{Generalization to hadron resonances}

The generalization of the weak-binding relation for unstable states is derived by adding a two-body decay channel in the effective field theory. We consider the same setup as before, and call the original two-body continuum channel 1. The threshold energy of the decay channel is set at $E=-\nu$ with $\nu>0$ measured from the threshold of channel 1. This changes the stable bound state to an unstable quasi-bound state with finite decay width. Representing the wave function of the quasi-bound state as $\ket{h}$, the compositeness $X$ is given by
\begin{align}
   \ket{h} = \sqrt{X}\ket{{\rm channel}\ 1} + \sqrt{1-X}\ket{\rm others} .
\end{align}
Note that $X$ represents the composite component of channel 1, and anything else are represented by $\ket{\rm others}$. For instance, the two-body component of the decay channel is included in $\ket{\rm others}$. In addition, because of the unstable nature of $\ket{h}$, the compositeness defined above becomes in general complex. Following the similar procedure, we can derive the weak-binding relation for the quasi-bound state $\ket{h}$ as
\begin{align}
   a_0=R\left\{\frac{2X}{1+X}
   +\mathcal{O}\left(\left|\tfrac{R_{\rm typ}}{R}\right|\right)
   +\mathcal{O}\left(\left|\tfrac{\ell}{R}\right|^3\right)\right\}
   \label{eq:weakbindingres}
\end{align}
with
\begin{align}
   R=\frac{1}{\sqrt{-2\mu E_{h}}},\quad
   \ell= \frac{1}{\sqrt{2\mu\nu}}
\end{align}
Several comments are in order. First, $a_{0}$ is the scattering length of channel~1, which is in general complex. Because the eigenenergy $E_{h}$ is complex, the ``radius'' $R$ is also a complex number. $\ell$ is the length scale associated with the threshold energy deference $\nu$. The first two terms in Eq.~\eqref{eq:weakbindingres} are essentially the same with Eq.~\eqref{eq:weakbinding}, and the last term can be neglected when the decay channel is located far away from the threshold of channel 1. For a fixed $R_{\rm typ}$ and $\ell$, the correction terms becomes small if the magnitude of the eigenenergy $|E_{h}|$ is decreased. In this case, the compositeness $X$ of the unstable state can be determined once the eigenenergy $E_{h}$ and the scattering length $a_{0}$ are given.

\subsection{Application to $\Lambda(1405)$}

Finally, we show an application of the generalized weak-binding relation in hadron physics. Here we consider $\Lambda(1405)$ resonance, for which two eigenstate poles are associated.\cite{Tanabashi:2018oca,Hyodo:2011ur} Here we consider the high-mass pole, whose energy is located close to the $\bar{K}N$ threshold. We collect the eigenenergy $E_{h}$ measured from the $\bar{K}N$ threshold and the $\bar{K}N$ scattering length $a_{0}$ in the analyses adopted in PDG\cite{Tanabashi:2018oca} in Table~\ref{tbl:L1405}. Applying the weak-binding relation~\eqref{eq:weakbindingres}, we determine the $\bar{K}N$ compositeness $X_{\bar{K}N}$. For the interpretation, we evaluate the real-valued compositeness $\tilde{X}=(1-|1-X|-|X|)/2$ with the estimated uncertainties from the correction terms (see Refs.~\citenum{Kamiya:2015aea,Kamiya:2016oao} for details). In all cases, the evaluated $\bar{K}N$ compositeness is close to unity. This indicates that the high-mass pole of the $\Lambda(1405)$ is dominated by the $\bar{K}N$ component.

\begin{table}
\tbl{Compositeness of the high-mass pole of $\Lambda(1405)$.}
{\begin{tabular}{@{}lcccc@{}}\toprule
Set & $E_{h}$ [MeV] & $a_{0}$ [fm] & $X_{\bar{K}N}$ & $\tilde{X}_{\bar{K}N}$ \\
\colrule
Set 1~(Refs.\citenum{Ikeda:2011pi,Ikeda:2012au}) & $-10-i26$ & $1.39-i0.85$ & $1.2+i0.1$ & $1.0^{+0.0}_{-0.4}$ \\
Set 2~(Ref.\citenum{Guo:2012vv}) & $-13-i20$ & $1.30-i0.85$ & $0.9-i0.2$ & $0.9^{+0.1}_{-0.4}$ \\
Set 3~(Ref.\citenum{Mai:2014xna})  & $\phantom{-0}2-i10$ & $1.21-i1.47$ & $0.6+i0.0$ & $0.6^{+0.3}_{-0.1}$  \\
Set 4~(Ref.\citenum{Mai:2014xna})  & $-\phantom{0}3-i12$ & $1.52-i1.85$ & $1.0+i0.5$ & $0.8^{+0.2}_{-0.2}$ \\
\botrule
\end{tabular}
}
\label{tbl:L1405}
\end{table}

\section{Summary}

We have discussed that the investigation of the structure of hadrons contains several subtleties, such as the treatment of unstable resonances and the ambiguity of the model space. The weak-binding relation and its generalization to unstable states can determine the structure of near-threshold $s$-wave states, avoiding the latter problem. At present, there is still no consensus on the characterization of the structure of unstable states. It will therefore be an important subject in future hadron spectroscopy to establish the prescription to characterize the structure of unstable states.

\section*{Acknowledgments}

The author is grateful to Feng-Kun Guo for kind invitation to Hadron 2019 conference. This work is supported in part by the Grants-in-Aid for Scientific Research
from JSPS (Nos. 19H05150 
and 16K17694). 



\end{document}